\documentclass[a4paper,11pt]{article}

\usepackage[utf8]{inputenc}
\usepackage[T1]{fontenc}  

\usepackage{graphicx}
\usepackage{amssymb}
\usepackage{mathrsfs}
\usepackage{amsfonts}
\usepackage{amsmath}
\usepackage{textcomp}
\usepackage{epsfig}
\usepackage{subfigure}
\usepackage{endnotes}   
\let\footnote = \endnote
\usepackage{blindtext}
\usepackage{makeidx} 

\usepackage{sidecap}
\usepackage{wrapfig}

\usepackage[table]{xcolor}  
\usepackage{caption}

\usepackage{nomencl}
\usepackage{ifthen}
\usepackage{lipsum}

\makenomenclature

\renewcommand{\nomgroup}[1]
{%
\ifthenelse{\equal{#1}{A}}%
{\item[]\hspace*{-\leftmargin}%
{\textbf{Symboles romains}}}%
{%
\ifthenelse{\equal{#1}{B}}%
{\vspace{3\parsep}\item[]\hspace*{-\leftmargin}%
{\textbf{Symboles grecs}}}%
{%
\ifthenelse{\equal{#1}{C}}%
{\vspace{3\parsep}\item[]\hspace*{-\leftmargin}%
{\textbf{Abréviations}}}%
}
}
}

\usepackage{hyperref}      
\hypersetup{
pdfpagemode=UseOutlines,      
pdfstartview=Fit,             
pdffitwindow=true,            
pdfpagelayout=TwoColumnsRight,
pdftoolbar=true,              
pdfmenubar=true,              
bookmarksopen=false,          
bookmarksnumbered=true,       
colorlinks=true,              
pdfauthor={Ton nom},     
pdftitle={Titre PDF},    
pdfcreator=PDFLaTeX,          %
pdfproducer=PDFLaTeX,         %
linkcolor=blue,               
urlcolor=blue,                
anchorcolor=black,            
citecolor=green,              
frenchlinks=true,             
pdfborder={0 0 0}             
}

\usepackage{multicol}    

\usepackage{amsmath} 
\usepackage{amsthm} 

\usepackage{multirow}
\usepackage{multicol} 
\usepackage[dvips,letterpaper,margin=0.60in,bottom=0.75in]{geometry}

\let\baraccent=\= 
\renewcommand{\=}[1]{\stackrel{#1}{=}} 

\theoremstyle{definition}

\theoremstyle{remark}

\usepackage{authblk}

\title{\textbf{Mitigating the photocurrent persistence of single ZnO nanowires for low noise photodetection applications}}

\author[1]{J-Ph. Girard}
\author[2]{L. Giraudet}
\author[1]{S. Kostcheev}
\author[2]{B. Bercu}
\author[3]{T.J. Puchtler}
\author[3]{R.A. Taylor}
\author[1]{C. Couteau}
\affil[1]{Light, Nanomaterials, Nanotechnologies (L2n), ICD-CNRS UMR 6281, Université de Technologie de Troyes, 10010 Troyes, France}
\affil[2]{Laboratoire de Recherche en Nanosciences, Université de Reims Champagne-Ardenne, 51685 Reims, France}
\affil[3]{Department of Physics, University of Oxford, Parks Road, Oxford OX1 3PU, United Kingdom}

\date{}

\begin{document}

 \maketitle

\begin{abstract}
In this work, we investigate the optoelectronic properties of zinc oxide (ZnO) nanowires, which are good candidates for applications based on integrated optics. Single ZnO nanowire photodetectors were fabricated with ohmic contacts. By taking current transient measurements in different atmospheres (oxygen, air, vacuum and argon), and at various temperatures, we point out the importance of surface effects on the electrical behaviour. Results confirm that oxygen chemisorption is responsible for the existence of a high photoconductive gain in these devices, and for the first time a two step process in the photocurrent rise transient is reported. A maximum gain of $G=7.8 \times 10^{7}$ is achieved. However, under certain conditions, the persistence of the photocurrent can last up to several hours and as such may prevent the device from operating at useful rates. From a knowledge of the photocurrent response mechanisms, we establish a method to restore the photodetector to its initial state, with very low dark current, by applying an appropriate gate voltage sequence. This advances the state of the art for these detectors towards commercial applications. 

\end{abstract}

\section{Introduction}

Photodetectors are one of the main devices (together with logical gates and waveguides) that are essential for applications in integrated optics. Over the last 20 years, improvements in nanofabrication techniques have enabled the creation of new photodetector devices based on nanobelts, nanowires and nanotubes \cite{intro_1D_ZnO}. These structures have different characteristics to those obtained from thin films or bulk materials. One dimensional (1D) metal oxide structures have been of great interest to the scientific community over recent years. Due to the high surface to volume ratio, these components benefit from properties specific to nanoscaling. ZnO nanowires have already proven their usefulness in the realisation of multiple components, such as field effect transistors \cite{FET_ZnO}, gas detectors \cite{biblio_gas_sensor}, photodetectors \cite{fast_response_ZnO_water}, LEDs \cite{biblio_LED} or even solar cells \cite{biblio_solar_cells}. \\
Zinc oxide is a wide band-gap semiconductor (3.37 eV) and as such, ZnO nanowires (nw) have been shown to be very promising UV detectors thanks to their significant photoconductive gain, as high as $10^{8}$ \cite{gain_ZnO}. This feature gives them sufficient sensitivity to detect small amounts of light (potentially at the single photon level).
Photoconduction is the physical effect that will interest us during this study. Under optical illumination, electrical charges are liberated in the semiconductor, increasing the density of free carriers and therefore decreasing the resistivity of the device. \\
So far, almost all of these detectors consist of an $n$ doped ZnO nw. ZnO nanowire photodetectors are known to rely on surface states, usually with the absorption of $\text{O}_2$ molecules at the surface, which contribute to the existence of gain \cite{prades_PCC_explaination}.
Unfortunately, this is accompanied by a persistence of the photocurrent (PPC) which can last for several hours, preventing the device from going back to its initial off-state after being illuminated. As a result, the device can not be used at high repetition rates. In this work, we present a complete study of ZnO based nw photodetectors. It combines electrical measurements at different pressures, temperatures, and gaseous compositions. A study of the nw photoluminescence in air and in vacuum was also performed. We addressed the photocurrent persistence and demonstrate a way to circumvent it. This step is based on a change in the polarity of the gate voltage, which facilitates the transit of charges to the electrodes and stems the persistence of the photocurrent.

\section{Experimental details}

\begin{figure}
    \centering
    \includegraphics[height=6cm]{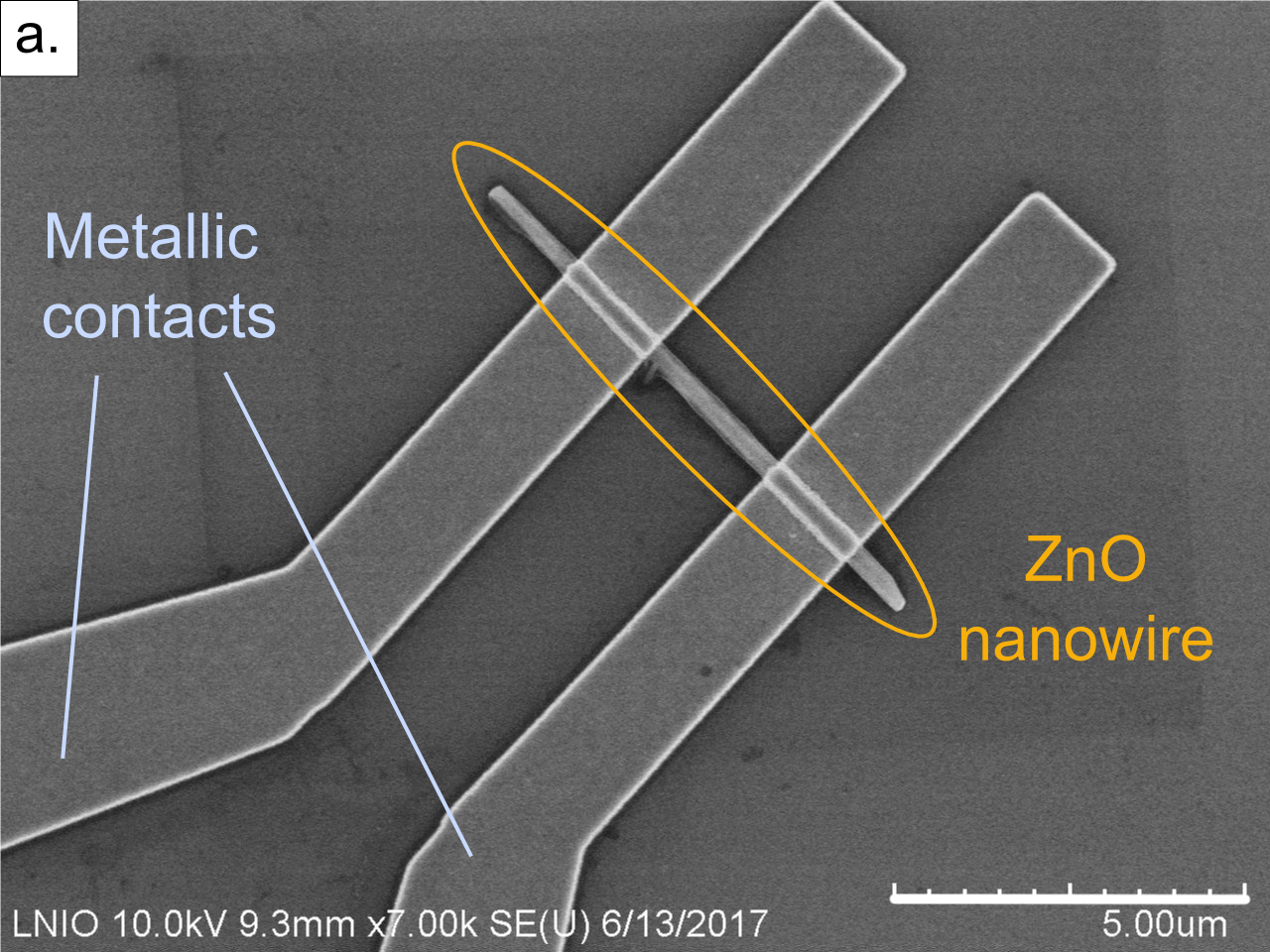}
    \includegraphics[height=6cm]{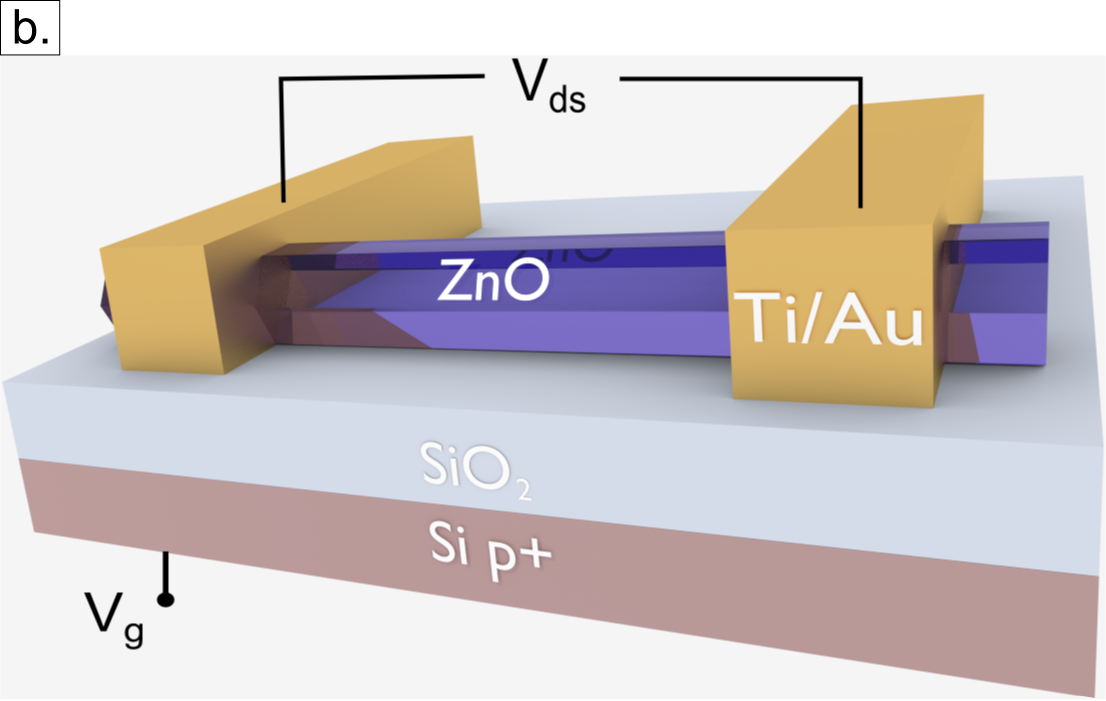}
    \caption{\textbf{a.} SEM image of a single nanowire photodetector. Scale bar is 5 $\mu$m. \textbf{b.} Schematic of the device, the drain-source bias $V_{\rm ds}$ is applied using the Ti/Au electrodes. A gate voltage $V_{\rm g}$ is applied below the sample, by polarising the silicon substrate.}
    \label{photo_PD19}
\end{figure}

Our ZnO nanowires were grown on a $c$-plane sapphire substrate by MOCVD (metalorganic chemical vapour deposition) in a horizontal reactor operating at a reduced pressure of $65\,  \rm mbar$. 
Oxygen and zinc were supplied by nitrous oxide ($8.6\, \text{mbar}$) and diethylzinc ($9.1\, \mu \text{bar}$) respectively, using heloium as a carrier gas. The hexagonal nanowires were grown at a temperature of $875$°C, and ended up forming a dense vertical forest of nanowires.
In order to create these photodetectors, ZnO nanowires were detached from the substrate by sonification in an ultrasonic bath with ethanol for $5\, \rm s$. Then, the nanowires in suspension were drop-casted on a silicon substrate ($p$ doped with an $\text{SiO}_2$ insulating layer of $300 \, \rm nm$ on top) \cite{carac_zno_iop}. \\
The electrical contacts ($\text{Ti/Au}$, of respectively $30\, \rm nm$ and $300\, \rm nm$) were fabricated using e-beam lithography on PMMA resist. For the alignement of the contacts to the nw, mechanical indents were drawn on the surface and then used as landmarks to localize the nanowires' positions \cite{Wei_Geng_fab}. Argon ion beam etching (IBE) was performed for $80\, \rm s$ at $5\, \rm mTorr$ and $50\, \rm W$ before the evaporation. The main purpose of this IBE step is to create oxygen vacancies in order to reduce an effective barrier at the metal/semiconductor interface. This step helps in the formation of good ohmic contacts, which are necessary for photoconduction with a gain higher than 1 \cite{ohmic_necessity}. Figure \ref{photo_PD19}a shows a SEM image of a ZnO nw contacted with Ti/Au electrodes. The average diameter of the nanowires is $250\, \rm{nm}$, their average length is $8\, \mu$m and the distance between the two electrodes is $2.5\, \mu$m. Figure \ref{photo_PD19}b is a schematic of the electrical contact setup. The gate voltage $V_{\rm g}$ is applied through the silicon substrate and the bias voltage $V_{\rm ds}$ is applied via the metallic electrodes. 

\section{Device Characterisation}

\begin{figure}
    \centering
    \includegraphics[height=6cm]{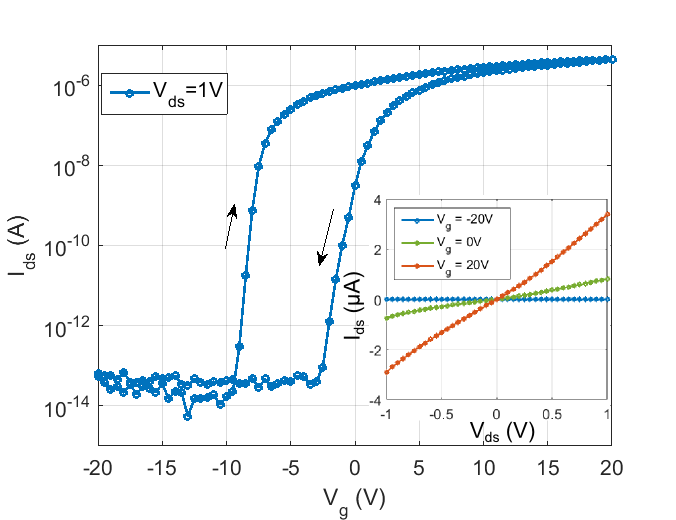}
    \caption{Room temperature (dark) $I_{\rm d}-V_{\rm g}$ transfer characteristic  obtained at $V_{\rm ds} =1 \rm V$. The measurements are performed in 1 bar oxygen atmosphere. The flow of electrons through the nanowire can be modulated by the gate voltage. Inset: $I_{\rm d}-V_{\rm ds}$ curves for -20, 0 and 20 V gate voltage showing an almost ohmic behaviour.}
    \label{IV_PD19}
\end{figure}

After nanofabrication, electrical properties were studied. Dark current measurements were performed at room temperature (approximately $300\,\rm K$ and $1\, \rm bar$) but in a pure O$_2$ atmosphere (see Figure \ref{IV_PD19}). In the transfer curve, we observe a drop of the dark current $I_{\rm d}$ over 7 decades below a gate voltage threshold around -$10\, \rm  V$, demonstrating a very low dark current capability and a very high ON/OFF ratio. Hysteresis is observed, the origin of which is attributed to carriers trapped at the nw surface, as discussed later. It is noteworthy that the hysteresis is highly dependent upon the exact experimental conditions (operating gas, pressure, speed of the voltage sweep). The nanowire behaves like a transistor with a threshold voltage around $V_{\rm g}= -10\,V$, below which the current no longer circulates in the nanowire.
The nanowire presents an almost linear behaviour for the $I_{\rm d}-V_{\rm ds}$ characteristic over a range of gate voltages $V_{\rm g} = [-20,20]\, \rm V$ (see inset Figure \ref{IV_PD19}). Using Ohm's law and a fit of these curves we can estimate the resistance of the device. 
The resistances are respectively $36.5\, \rm{T}\Omega$, $1.25\, \rm{M}\Omega$ and $296\, \rm{k}\Omega$ for $V_g = -20\,\rm{V}$, 0 and $20\, \rm V$. These values are consistent with our previous measurements \cite{Wei_Geng_fab}. 
The ZnO nanowire resistivity can also be estimated from $\rho = R \frac{\phi}{l_{\rm ds}}$, with $R$ the electrical resistance ($\Omega$), $\phi$ the nanowire cross section area ($\rm{m}^2$) and $l_{\rm ds}$ the distance between the two electrodes ($\rm m$). 
The average nanowire diameter $D$ was determined from SEM images to be around $250\, \rm nm$. The nanowires are homogeneous in diameter and the length of the device $l_{\rm ds}$ is $2.5\, \mu \rm m$ (see Figure \ref{photo_PD19}).
The resistivities found are $3\,10^{8}\, \Omega . \rm cm$, $9.8\, \Omega .\rm  cm$, $2.32\, \Omega . \rm cm$  for respectively $V_{\rm g}=-20\,\rm V$, $V_{\rm g}=0\, \rm V$ and $V_{\rm g}=20\, \rm V$.
In the literature a wide range of values can be found (from $10^{-3}$ to $10^5 \, \Omega . \rm cm$
\cite{difficult_nw_characterisation}). It is difficult to make a comparison though as in most reported results, the value of the gate voltage is not specified or even controlled during the experiment. 
Also, it is important to highlight that the contact resistances are not taken into account in these resistivity calculations. Consequently the nanowire resistivities discussed here are an overestimate. Nevertheless, $I_{\rm d}-V_{ \rm ds}$ measurements have been performed on a previous sample using a 4-point contact design (fabrication details can be found in \cite{these_wei}), where the contact resistance contribution can be separated from the intrinsic nanowire contribution. The intrinsic nanowire resistivity was then estimated to be around $\rho_{\rm nw}=0.32\, \Omega . \rm cm $ for $V_{\rm g}=0\, \rm V$. These values are consistent with the literature \cite{difficult_nw_characterisation}, where resistivities ranging from 2 to 12 $\Omega . \rm cm$ are found for similar nanowire diameters. To conclude, the contact resistances are the main contribution to our device total resistance. Therefore there is an opportunity to further improve the performance of our PD by reducing the contact resistances.

\begin{figure}
    \centering
    \includegraphics[height=6cm]{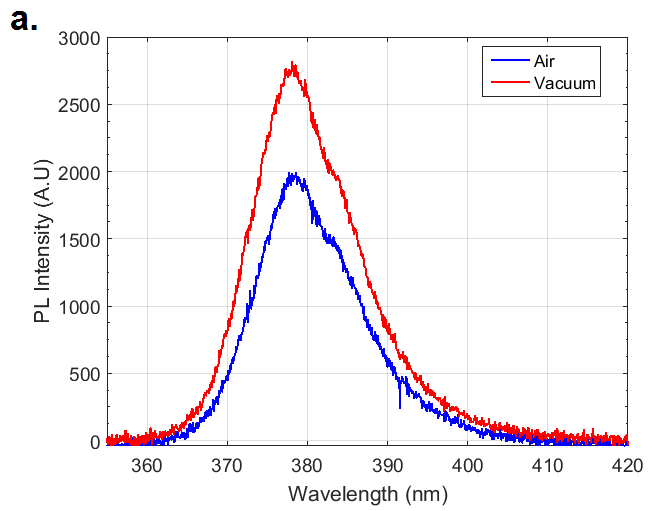}
    \includegraphics[height=6cm]{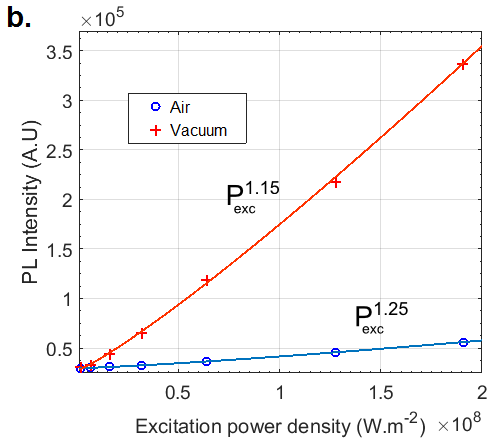}
    \caption{\textbf{a.} Photoluminescence spectra of ZnO nanowires under vacuum and ambient air, at room temperature for an incident optical power of $1.5 \times 10^{6} \, \rm{W.m}^{-2}$ \textbf{b.} Nanowire photoluminescence intensity variation versus excitation optical power density, in air and in vacuum. The fits are carried out with equation $I_{\rm PL}=a P_{\rm exc}^b+c$. }
    \label{comparaison_air_vide}
\end{figure}

Photoluminescence (PL) measurements were performed both at room temperature under vacuum and in ambient air. Measurements have been done using a frequency-tripled, mode-locked Ti:Sapphire laser as an excitation source ($\lambda = 266\, \rm nm$, pulse duration = $150\, \rm fs$, repetition rate = $76 \, \rm MHz$). The beam was focused on the sample with a $36\times$ reflecting objective (numerical aperture of 0.5), the photoluminescence was then collected through the same objective and directed to an Andor $30\, \rm cm$ Shamrock spectrometer (with a spectral resolution of $0.07\, \rm nm$) For an excitation at $266\, \rm nm$ (above the bandgap energy of ZnO), one can observe a luminescence maximum at $378\, \rm nm$. The position of this maximum is in agreement with the results for PL from ZnO \cite{PL_T_ambiante} ($383\, \rm nm$).
Figure \ref{comparaison_air_vide}a presents the PL intensity of a collection of ZnO nanowires for wavelengths between $\lambda = 350\, \rm nm$ and $\lambda = 420\, \rm nm$, in air and under vacuum. The PL shape is similar in air and under vacuum, however the intensity in vacuum is higher than in air, which indicates that non-radiative recombination is enhanced by the presence of $\text{O}_2$ (as discussed later). The PL intensity was also recorded at various excitation power densities (see Figure \ref{comparaison_air_vide}b). The fits were carried using the equation $I_{\rm PL}=a P_{\rm exc}^b+c$. The PL intensities do not increase strictly linearly with the excitation power density. This tends to indicate the presence of nonlinear effects, especially for the sample in air, since its curve follows $I_{\rm PL}^{\rm air} \propto P_{\rm exc}^{1.25}$ (compared to its counterpart in vacuum $I_{\rm PL}^{\rm vac} \propto P_{\rm exc}^{1.15}$). These nonlinear effects bypass the classical band to band recombinations and can imply bound-excitons, biexcitons, etc. \cite{example_PL_ZnO}.\\ 

The nanowire photocurrent time response has been investigated in various atmospheres in order to investigate the persistence of the photocurrent. The study of photoconduction in O$_2$ is relevant because it has been shown \cite{prades_PCC_explaination} that oxygen leads to a reduction in the PPC. Illumination was provided by various fiber pigtailed LEDs, synchronised with a Keysight E5270 precision I(V) Analyser. The nanowires were probed in a cryostat under various temperatures and atmospheres. When exposed to photons of energy greater than the gap of ZnO, the photoresponse evolution drastically changed depending on the surrounding gas and pressure. \\

\begin{figure}
    \centering
    \includegraphics[height=6cm]{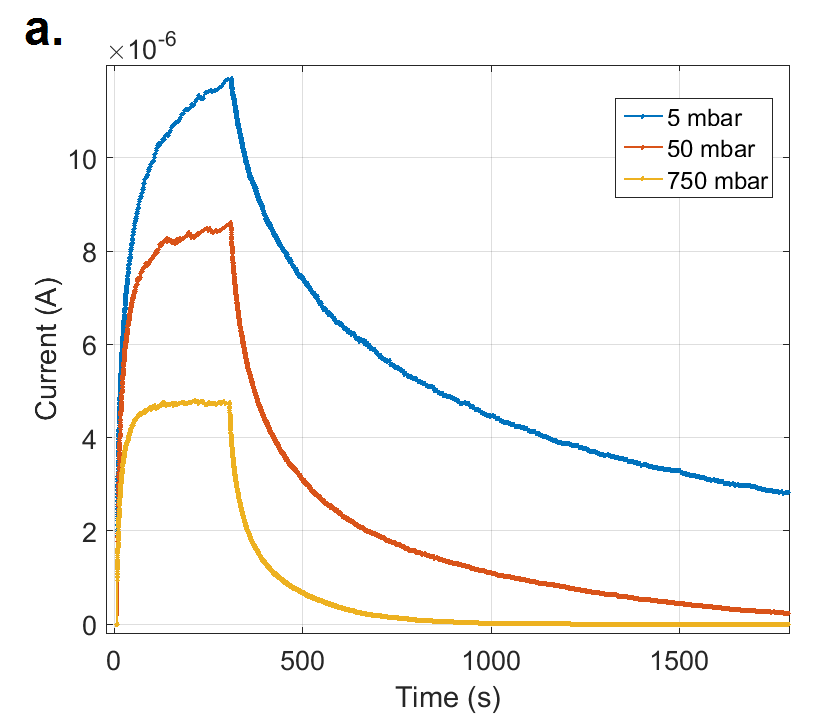} 
    \includegraphics[height=6cm]{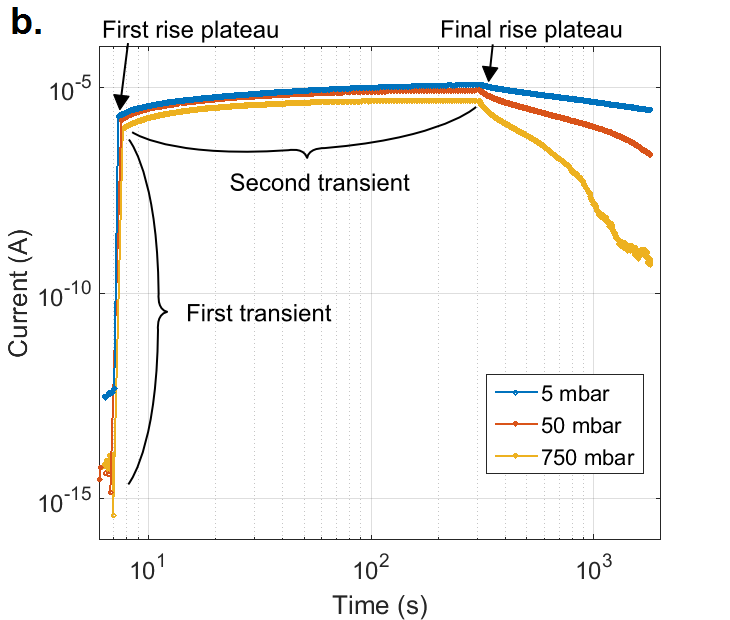}\\
    \includegraphics[height=5.2cm]{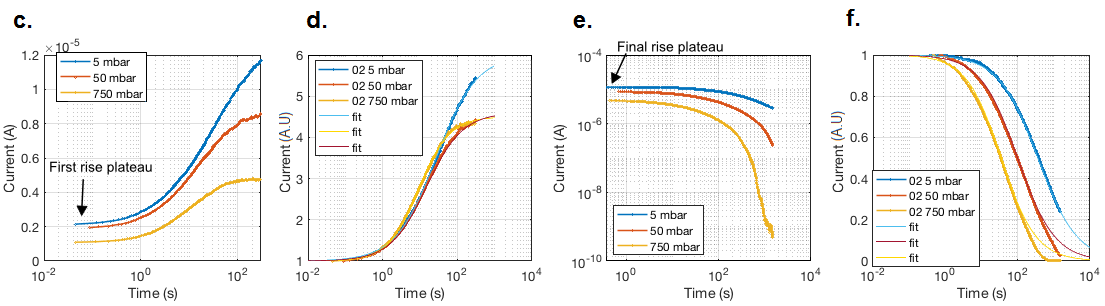}
    \caption{\textbf{a.} Nanowire photocurrent response under a pure oxygen atmosphere at $5\, \rm mbar$, $50\, \rm mbar$ and $750\, \rm mbar$ (respectively blue, red and orange lines). Illumination at $\lambda = 365\, \rm nm$ starts at t=0 s and stops at t=300 s.\textbf{b.} Same curve on a log scale. \textbf{c.} and \textbf{e.} Raw data for rise and fall transients respectively. \textbf{d.} and \textbf{f.} Normalized rise and fall transients respectively with their fitted curves.}
    \label{effet_oxygen}
\end{figure}

\begin{figure}
    \centering
    \includegraphics[height=6.7cm]{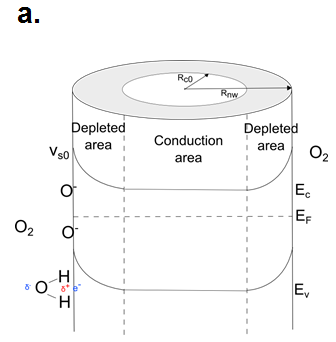}
    \includegraphics[height=6.8cm]{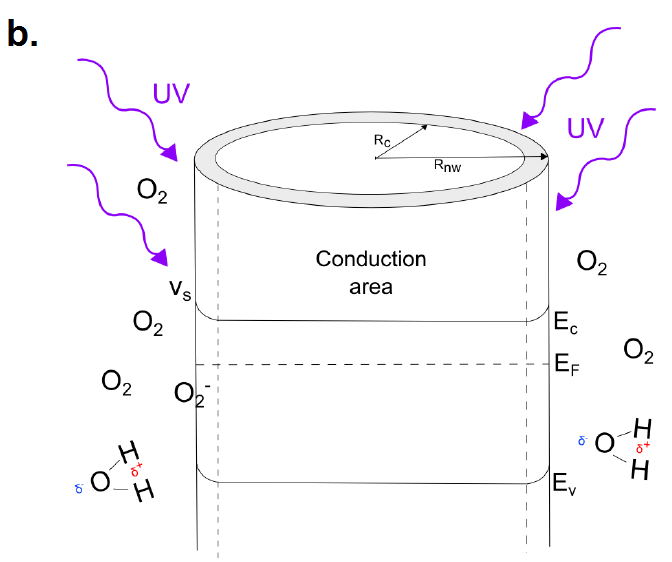}
    \caption{Schematic of the influence of oxygen on the electrical properties of the nanowire. \textbf{a.} In the dark, oxygen molecules are absorbed on the surface and create potential barrier which reduce the density of free electrons near the surface. \textbf{b.} UV light contributes to the desorption of molecules previously trapped, electrons are released and so electrical conductivity is enhanced.}
    \label{schema_photoconduction}
\end{figure}

The nanowire photoresponse was first studied in a pure oxygen atmosphere (see Figure \ref{effet_oxygen}a and b). These graphs represent the photocurrent intensity versus time. The measurement begins in the dark with a dark current below $10^{-13}\, \rm{A}$. In the interval $t \in [0, 300\, \rm{s}]$ the sample is illuminated with UV light at $ \lambda = 365 \, \rm nm $. At t=$300\, \rm s$ the illumination is stopped and the photocurrent decay is measured up to t=$1800\, \rm s$.
The photocurrent transients have been recorded under 3 oxygen pressures ($5\, \rm mbar$, $50\, \rm mbar$ and $750\, \rm mbar$). The transients (rise and decay of the photocurrent) are slow (much slower than a second) and vary with the oxygen pressure. The lower the pressure, the higher the photocurrent, and the longer its decay time. \\
The relatively slow response of the nanowire may seem to contradict with expectations. Indeed in photoconductors, the photocurrent gain is due to the relative transit times of electrons and holes \cite{fundamentals_photonics}. In order to maintain neutrality, the fastest carrier (here an electron) is injected into the photoconductor from one electrode and transits to the other electrode, again and again, as long as the slower photogenerated carrier (here a hole) does not reach its collecting electrode. Therefore several carriers transit through the biasing circuit for each single absorbed photon, generating a photoconduction gain. In our case where holes are much slower than electrons, the gain is given by $ G \approx \frac{t_{\rm{p-transit}}}{t_{\rm{n-transit}}} $. The transit time of the holes from one electrode to the other is $t_{\rm{p-transit}}$. The transit time of electrons from one electrode to the other can be approximated by $ t_{\rm {n-transit}} = \frac{L^2}{\mu_n V_{\rm ds}} \approx 0.3 \, \rm ns $ and the gain of the detector was measured as $ G \sim 10^6 $ (see later in the article). As a result the photocurrent should reach its maximum after about $ t_{\rm{p-transit}} = G \times t_{\rm{n-transit}} \approx 0.3 \, \rm ms $. Our measurements show a time response 3 orders of magnitude slower. Another process is obviously responsible for the observed photocurrent transients. 
A more detailed analysis of rise and fall transients appeared necessary, as illustrated Figure \ref{effet_oxygen}a-f where the raw and normalized rise and fall transients are presented. \\

 Considering the rise transients, we observed two distinct contributions to the photocurrent, each with its own characteristic time constant and saturation value. To the best of our knowledge, this is the first time observation of two distinct photocurrent responses has been reported and discussed. For the sake off analysis, the second rise transient and the photocurrent decay have been plotted on Figure \ref{effet_oxygen} c-f on a log scale and from their own time origin. \\
 
 A first fast transient occurs as soon as illumination starts, up to a first rise plateau. The fast rise has a response time of the order of the calculated one $ (0.3 \, \rm ms) $ and is too fast to be resolved using our equipment (limited to a few ms). It contributes almost completely to the value of the photocurrent gain. It is likely that this first and fast photo-response does not involve  reactions occurring at the surface of the nanowire. It is mainly due to the time required to reach an equilibrium of the free carrier concentrations, electrons and holes, under illumination. The carrier concentrations are determined by optical generation of photo-carrier pairs (electron-hole pairs), photo-carrier recombination, and drift-diffusion of electrons and holes towards the contacts. This first rise time depends on factors such as the optical power, the recombination rates, and the carrier mobilities, and is too short for our measurement setup. This is why only a first plateau is observed, with only a slight dependence on the oxygen pressure. \\

The second rise transient, which we will call secondary photocurrent evolution, is shown in Figures \ref{effet_oxygen}b and c. The secondary photocurrent contributes only weakly to the gain (only by a factor of about 5), and can take a very long time to reach a plateau. At low oxygen pressure $(5\, \rm mbar)$ the second plateau is not reached after $300\, \rm s$.
The evolution of the secondary photocurrent rise transient can be approximated by: 

    \begin{equation}
 I = I_0 \left( 1+\frac{1}{\frac{1}{\left(t/{\tau}_{\rm {rise}}\right)^{\alpha}}+1/A}   \right)      
 \label{eq_1}
    \end{equation}

where $ I_0 $ is the first (fast) photocurrent plateau, $ A = \frac{I_{\infty}}{I_0} -1 $ accounts for the maximum increase of the secondary photocurrent, and ${\tau}_{\rm rise}$ is the characteristic time of the second rise transient. Fitting the secondary rise transient using this expression gives the various constants given in Table \ref{rise_fall_time_table}.
For all transients $ \alpha $ is close to unity. Therefore, at the very beginning of this process and before saturation occurs, the secondary rise transients increase almost linearly with time. The time constant ${\tau}_{\rm rise}$ of the process responsible for this transient is of the order of 3 seconds, and is almost independent of the oxygen pressure. The saturation value of the photocurrent depends slightly on the oxygen pressure. The secondary gain $I_{\infty}/I_0$ is about 5. In comparison the gain of the fast first photocurrent transient ranges from 6 to $ 12\times 10^4 $ for this experiment (see Table \ref{rise_fall_time_table}a), contributing to most of the total gain. \\

The second transient is attributed to the modification of the nanowire surface. In the presence of oxygen molecules and under dark conditions, oxygen can be chemisorbed at the nanowire surface in the form ${\rm O}_{2}^{-}$ \cite{comini}.
As illustrated in Figure \ref{schema_photoconduction}, these molecules are responsible for a surface depletion, the free electron density $n$ is reduced to ensure neutrality of the nanowire, and consequently the nanowire conductivity decreases. Indeed, in dark conditions, neglecting surface level pinning, neutrality requires the free electron density $n$ to compensate exactly for the residual donor concentration: ${N}_{D}^{+}=n$. In the presence of chemisorbed ${\rm O}_{2}^{-}$ molecules, the averaged free electron concentration is reduced to maintain global neutrality following: $n = {N}_{D}^{+} - S_{O_{2}}\frac{4}{D}$ where $S_{O_{2}}$ is the surface density of chemisorbed ${\rm O}_{2}^{-}$  molecules and $\frac{4}{D}$ a coefficient to account for the surface to volume ratio of a nanowire slice. The free electrons are pushed away from the surface due to the negative charge of the ${\rm O}_{2}^{-}$ molecules, which creates a depletion region all around the nanowire surface. The nanowire conductivity is therefore reduced in the presence of chemisorbed ${\rm O}_{2}^{-}$ molecules. The application of a negative gate voltage further reduces the free electron concentration to a point where the dark current becomes lower than the measurement noise, well below $0.1 pA$. When illuminated, photogenerated holes are attracted to the surface, where they recombine with the excess electrons of the chemisorbed ${\rm O}_{2}^{-}$ molecules, contributing to the oxygen molecule desorption (in a neutral form). The surface depletion disappears and the nanowire conductivity increases, rising the photocurrent. This loss of chemisorbed oxygen can be partly counterbalanced by the chemisorption of new oxygen molecules, if available in the ambient atmosphere. The surface density of adsorbed oxygen molecules will reach an equilibrium due to these two competing processes. The lower the oxygen pressure, the deeper the surface oxygen desorption, the higher the electron density, the higher the conductivity, and the higher the secondary photocurrent plateau. The speed of the ${\rm O}_{2}^{-}$ desorption process does not depend on the surrounding oxygen pressure within the experimental range. A high surrounding oxygen pressure helps reaching low dark current conditions, however a low oxygen pressure ensures deeper ${\rm O}_{2}^{-}$ desorption and higher photocurrent gain. \\

 An examination of the photocurrent decay after illumination has been turned off reveals a clear dependence of the fall transients on the oxygen pressure: the higher the pressure the faster the photocurrent decay. The transient can be approximated by the following expression:
\begin{equation}
I = I_{\infty} \left( \frac{1}{\left( t/{\tau}_{\rm fall}\right) ^{\alpha}+1}   \right) 
\label{eq_2}
\end{equation}

The parameters obtained for the three oxygen pressures are reported in Table \ref{rise_fall_time_table}b. Again $ \alpha $ is close to unity for all pressure conditions, indicating a photocurrent decaying closely to $ 1/t $. The decay process time constant ${\tau}_{\rm fall}$ is however very sensitive to the oxygen pressure, as already observed in the literature \cite{absorption_O2_pression}. The time constant ${\tau}_{\rm fall}$  increases from $ 35\, \rm s $ to $ 360\, \rm s $ when reducing the oxygen pressure from $(750\, \rm mbar)$ to $(5\, \rm mbar)$. The decay process involves a regeneration of the ZnO surface with fresh oxygen. When fresh oxygen is lacking due to external conditions the regeneration process becomes very slow or can even stop. \\

\begin{table}[]
\centering
\begin{tabular}{c|c|c|c|c|cccccl}
\cline{2-5} \cline{7-10}
$\textbf{a.}$ & $\textbf{Rise}$                                  & 5 mbar          & 50 mbar         & 750 mbar        & \multicolumn{1}{c|}{$\textbf{b.}$} & \multicolumn{1}{c|}{$\textbf{Fall}$}   & \multicolumn{1}{c|}{5 mbar} & \multicolumn{1}{c|}{50 mbar} & \multicolumn{1}{c|}{750 mbar} &  \\ \cline{2-5} \cline{7-10}
              & $\alpha$                                         & 0.8             & 1.05            & 0.9             & \multicolumn{1}{c|}{}              & \multicolumn{1}{c|}{$\alpha$}          & \multicolumn{1}{c|}{0.8}    & \multicolumn{1}{c|}{0.85}    & \multicolumn{1}{c|}{0.9}      &  \\ \cline{2-5} \cline{7-10}
              & $\tau_{\rm rise}$ (s)                                & 3.6             & 2.8             & 3.6             & \multicolumn{1}{c|}{}              & \multicolumn{1}{c|}{$\tau_{\rm fall}$ (s)} & \multicolumn{1}{c|}{360}    & \multicolumn{1}{c|}{100}     & \multicolumn{1}{c|}{35}       &  \\ \cline{2-5} \cline{7-10}
              & $I_{\infty}/I_0$, $2^{\text{nd}}$ transient gain & 6               & 4.5             & 4.6             &                                    &                                        &                             &                              &                               &  \\ \cline{2-5}
              & $1^{\text{st}}$ transient gain                   & $1.2\times 10^{5}$ & $1.1\times 10^{5}$ & $6.2\times 10^{4}$ &                                    &                                        &                             &                              &                               &  \\ \cline{2-5}
                            & Total gain $G$                   & $7.2\times 10^{5}$ & $5.0\times 10^{5}$ & $2.9\times 10^{5}$ &                                    &                                        &                             &                              &                               &  \\ \cline{2-5}
\end{tabular}
\caption{\textbf{a.} Extracted constants of rise \textbf{a.} and fall \textbf{b.} secondary photocurrent transients. All transients $\alpha$ values are close to 1, indicating an almost linear time dependence of all processes. The various rise constants are almost independent of the oxygen pressure. Conversely, the fall transient is faster by a factor of 10 when increasing of the oxygen pressure from $ 5 \, \rm mbar $ to $ 750 \, \rm mbar $. The photocurrent gains have been estimated using Equation \eqref{eq_Gain} as described in the text.}
\label{rise_fall_time_table}
\end{table}

\begin{figure}
    \centering
    \includegraphics[height=6cm]{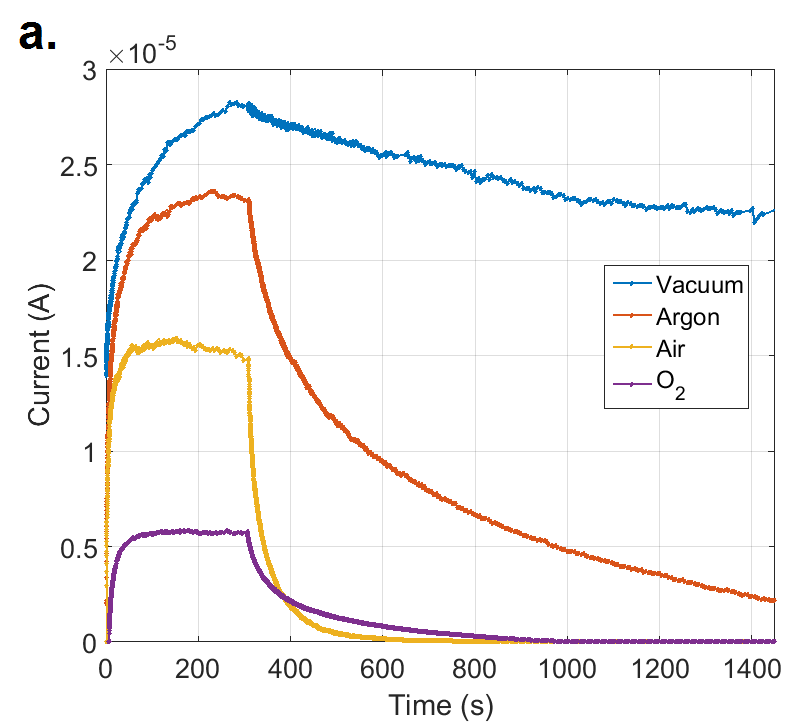}
    \includegraphics[height=6cm]{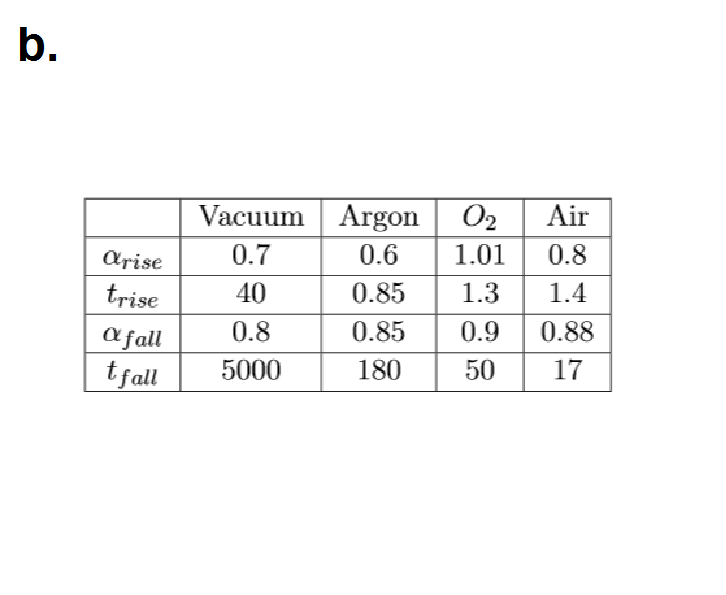}
    \caption{\textbf{a.} Photocurrent versus time under different atmospheres at room temperature. Illumination at $365\, \rm nm$ starts at $t=0\, \rm s$ and stops at $t=300\, \rm s$. The photocurrent persistence is much longer under vacuum and argon than in presence of oxygen. \textbf{b.} Values obtained from the fits of the rise and fall transients of the photocurrent using the formula \eqref{eq_1} and \eqref{eq_2} described in the text. }
    \label{I_ph_effet_atm}
\end{figure}

Figure \ref{I_ph_effet_atm} presents similar experiments but for different gases at room pressure (argon, air, $\rm{O}_2$) and under vacuum (approximately $ 10^{-4}\, \rm mbar $). The initial dark currents depend on the atmosphere. The dark currents in air and oxygen are below $10^{-13} \, \rm{A}$, and are higher in vacuum and argon (respectively $14 \, \mu \rm{A}$ and $2 \, \mu \rm{A}$).  
Data have been analysed using the methodology previously described. The transient phases (increase and decrease of the photocurrent $I_{\rm ph}$) appear rapid in the presence of oxygen and air, and the decay is even faster in air, suggesting that humidity plays some role in the oxygen chemisorption process, increasing the chemisorption rate. An increase of a factor of 100 is observed between the decay times in oxygen and vacuum. This confirms that the persistence of the photocurrent can be reduced in the presence of oxygen. For the rise processes, analysis show that similar time constants are obtained at ambient pressure regardless of the atmosphere, but the process seems slower in vacuum, unexpectedly. We believe that the extracted time constant in vacuum is not accurate since the photocurrent is already very high when the illumination starts. In vacuum the surface hardly regenerates its oxygen density and the photocurrent almost never returns to a comparable initial state than in other experiments. Actually in the present experiment the dark current under vacuum was about $ 14\, \mu \rm A $, about half of the saturation photocurrent.\\

From these various experiments the rise and fall transients of the photocurrents can be well understood. Ambient $\text{O}_2$ molecules are physisorbed on the surface of the nanowire, usually because of surface defects such as oxygen vacancies \cite{origin_doping_ZnO}, as observed in previous works \cite{prades_PCC_explaination}. Once physisorbed, oxygen molecules can trap electrons, becoming chemisorbed in the form $\rm{O}_{2}^{-}$ \cite{comini}, which increases the band bending at the surface to ensure neutrality. The band bending involves ionised donors, the presence of which is confirmed by conductivity measurements under various gate voltages, as shown above. The surface depletion shrinks the conductive channel (see Figure \ref{schema_photoconduction}) and reduces the number of free electrons. The nanowire conductivity tends to be reduced by the presence of chemisorbed oxygen at the nanowire surface. This is an efficient effect in nanowires because of their high surface to volume ratio. A similar process can arise from dipolar $\text{H}_2\text{O}$ molecules, which in the presence of $\text{O}_2$ can react to form two $\text{OH}^{-}$ groups \cite{TiO2_absorption_eau}. This might be one explanation for the very fast recovery process of the dark photocurrent in ambient air.\\

\begin{figure}
    \centering
    \includegraphics[height=5.4cm]{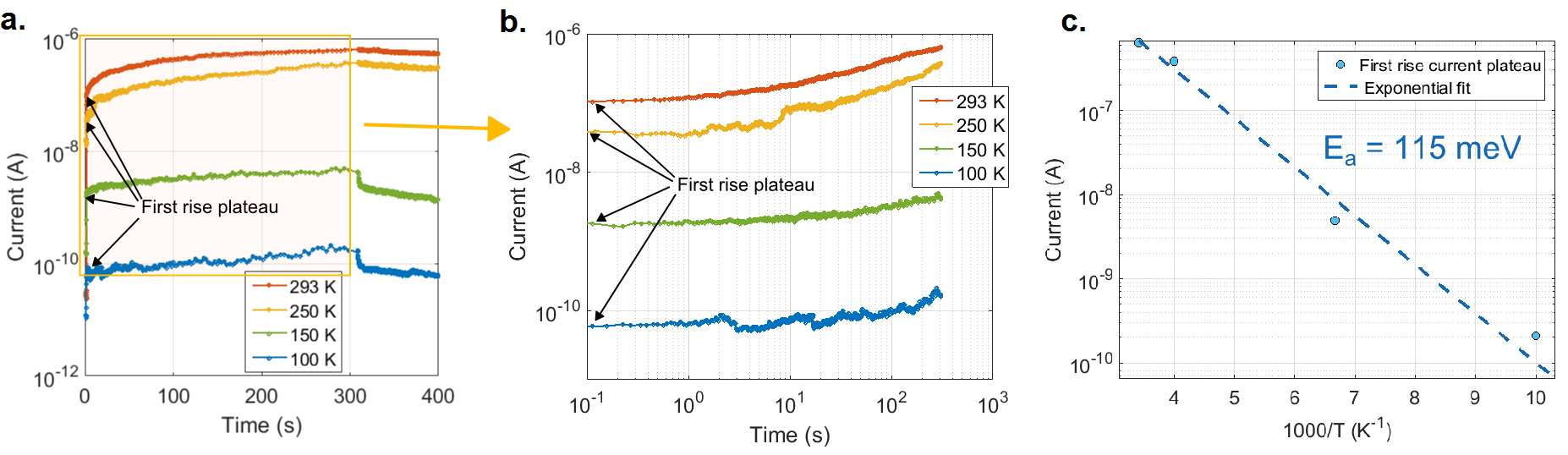}
    \caption{\textbf{a.} Photocurrent measurements at different temperatures ($100\,\rm K$-$293\, \rm K$), in oxygen atmosphere ($0.5\, \rm mbar$), illuminated at $365\, \rm nm$. First photocurrent rise plateau is indicated. \textbf{b.} Data for rise transient on a log scale. \textbf{c.} First photocurrent plateau versus $1000/T$: activation energy is determined from the formula $I_{\rm ph} = I_0 \exp{ \left( -E_{\rm a}/k_BT \right)}$.}
    \label{richardson}
\end{figure}

 Temperature measurements have been performed to better characterize the photoconduction mechanism.  The photocurrent response has been measured at $V_{\rm g} =-20\, \rm V$ (in order to reduce dark current) for temperatures ranging from $100\,\rm K$ to $293\, \rm K$, under low oxygen pressure of $0.5\, \rm mbar$. As previously we observed two successive increases of the photocurrent (see Figure \ref{richardson}a and b). A first plateau is reached almost instantly, then the current slowly increases towards a second plateau. Unfortunately the second plateau is not reached after $(300\, \rm{s})$ under a low oxygen pressure $(0.5\, \rm mbar)$. This can be understood since the second plateau involves oxygen desorption at the nanowire surface. The activation energy of this process is high, about $(0.7\, \rm eV)$ as determined by \cite{chemisorption_O2_ZnO}, therefore the process becomes very slow at low temperatures, as confirmed by our experiments. However, as shown Figure \ref{richardson}b, the first plateau is clearly reached and could be analysed. From a plot of the photocurrent plateau $I_{\rm ph}$ versus $1000/T$ an activation energy of $E_a = 115\, \rm meV$ is determined (see Figure \ref{richardson}c).
 
 This activation energy could be a signature of the surface equilibrium reached under dark conditions. The surface depletion at the nanowire surface, due to chemisorbed oxygen molecules, is responsible for the separation of photo-generated electrons and holes. The depletion width reduces the electron-hole recombination rate, and consequently influences the equilibrium concentrations of free electrons and holes. This equilibrium is responsible for the high photocurrent gain of nanowires compared to bulk photoconductors. The detailed analysis of the equilibrium reached in dark conditions, such as the surface depletion width and the chemisorbed oxygen surface density, is complex and beyond the scope of this paper. It would probably require complementary experiments. However the present activation energy determination could help in understanding the surface dynamics of ZnO nanowires.\\

\begin{figure}
    \centering
    \includegraphics[height=6.5cm]{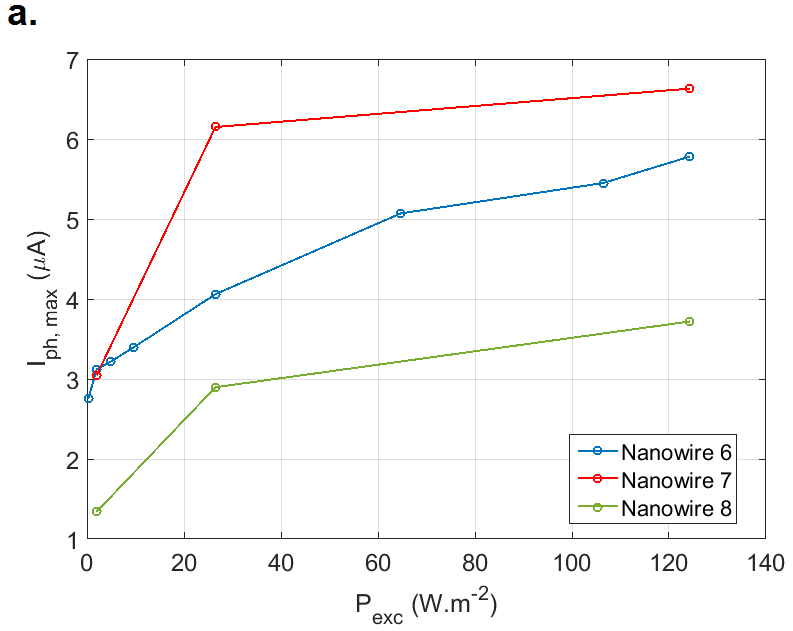}
    \includegraphics[height=6.5cm]{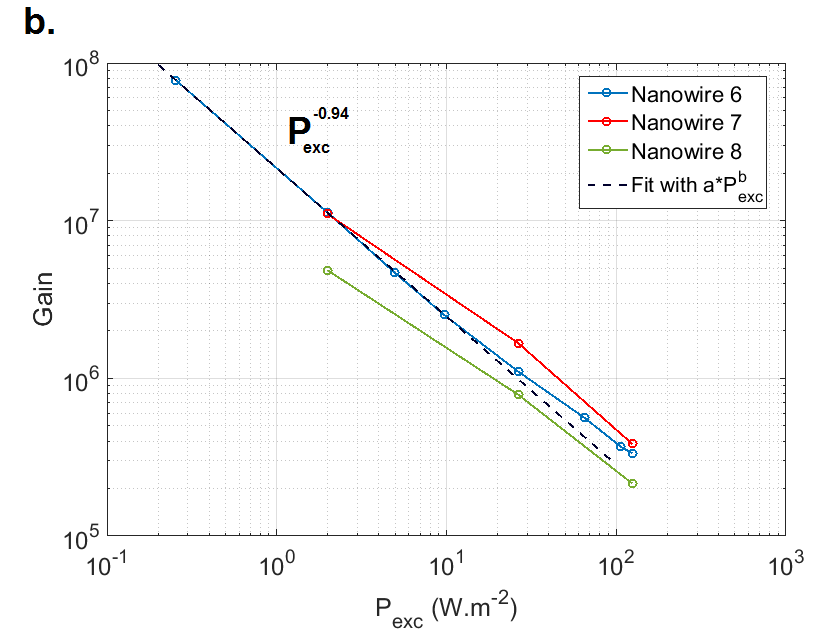}
    \caption{\textbf{a.} Maximum photocurrent at $293\, \rm K$ in 1bar $\rm{O}_2$ atmosphere, as a function of light excitation power density for three different nanowires. \textbf{b.} Nanowire photodetector gain as a function of the excitation power density for the three nanowires.}
    \label{I_ph_vs_P}
\end{figure}

In order to gather more information about the photoconduction in our component, the photoresponse has been studied as a function of the incident optical power density in an oxygen atmosphere. In Figure \ref{I_ph_vs_P}a, the maximum photocurrent is presented as a function of the incident optical power density for three nanowires. The maximum photocurrents are within the same order of magnitude for the three nanowires studied, and increase with incident optical power as expected. The photocurrent gain can be estimated from the following expression:
\begin{equation}
 G = \frac{I_{\rm ph}/q }{\eta _{\rm ext} \, P_{\rm exc}/(h \nu)}  
 \label{eq_Gain}
 \end{equation}
where $q$ is the elementary charge, $\eta _{\rm ext}$ is the external quantum efficiency, $I_{\rm ph}$ is the photocurrent, $P_{\rm exc}$ is the incident optical power, and $h \nu$ is the photon energy. In a first approximation $\eta _{\rm ext}$ can be estimated considering a rectangular nanowire with a width and a thickness of $d_{\rm nw}=250\, \rm nm$ and a length $L=2.5\, \mu \rm m$. Photon absorption in the ZnO nanowire is estimated using the Beer-Lambert equation, with an absorption coefficient $ \alpha = 3 \times 10^{3}\,  \rm{cm}^{- 1} $ as measured previously \cite{Wei_absorption} for a nanowire of similar diameter. The calculated absorption is almost complete ($>92\%$). The reflection at the air/nanowire interface is approximated to $18\%$, as it would be for a plane wave under normal incidence. Finally, taking into account the illumination setup ($50\, \mu \rm m$ diameter cleaved multimode fiber) and the nanowire surface, the external quantum efficiency is evaluated to be  $\eta_{\rm ext} = 2.4\times 10^{-4}$. \\

From the measured photocurrent $I_{\rm ph}$, the gain can be calculated. Results are shown in Figure \ref{I_ph_vs_P}b. One notices for all the nanowires a similar decrease of the gain when increasing the incident optical power. The gain variation can be fitted by an inverse power law  $G = P^{-k}$, with $k=0.94$. The photocurrent gain is therefore almost inversely proportional to the incident optical power. This is in agreement with Soci \textit{et al.} \cite{gain_ZnO} where $k$ was found to be $0.7$. According to this work, the photocurrent can be expressed from the photon absorption rate F ($\rm{s}^{-1}$) as:
\begin{equation} 
I_{\rm ph} = q \frac{\tau_l^{0}}{\tau_{t}} \frac{F}{1+(F/F_0)^k} 
\label{eq_4}
\end{equation}
where $q$ is the elementary charge, $\tau_t$ the electron transit time, $\tau_l^0$ is the hole transit time (or the electron lifetime), and $F_0$ is a constant. The photon absorption rate is directly proportional to the incident optical power: $F \propto P_{\rm exc} $. Consequently the photocurrent varies like $I_{\rm ph} \propto P_{\rm exc}^{1-k} $ with the incident optical power. This is a counter-intuitive result when $k$ is close to 1, as $I_{\rm ph}$ becomes almost independent of the incident optical beam power. Our device is in this situation: as can be seen Figure \ref{I_ph_vs_P}a, the photocurrent is only multiplied by a factor 2 when the incident optical power $P_{\rm exc}$ is increased by more than 450. It means the sensitivity is increased at low optical power, and therefore, the device is suited for low intensity light detection.\\

\begin{figure}
    \centering
    \includegraphics[height=6cm]{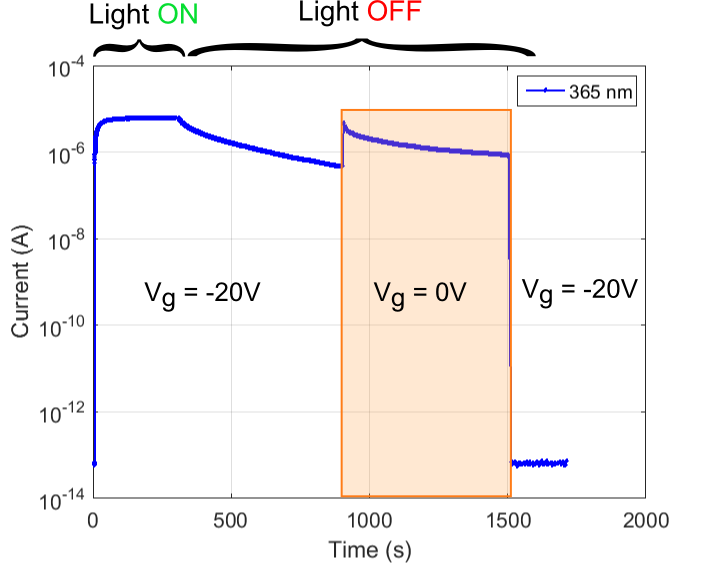}
    \caption{Nanowire response to illumination at $365\,$nm in an oxygen atmosphere and at room temperature (note the log scale). Return to the initial off-state is accelerated using an appropriate gate pulse.}
    \label{I_ph_vs_t}
\end{figure}

The persistence of the photocurrent is detrimental to many applications of the device since it prevents high frequency operation. We discuss now a method to quickly restore the nanowire photodetector to its initial state once illumination has been turned off. As shown Figure \ref{I_ph_vs_t} (note the log scale), the photocurrent decay can be very long after illumination is turned off in normal conditions: it may take several hours. This prevents using the photodetector at useful rates. Luckily, an appropriate gate voltage pulse happens to stop the persistent photoconductivity and restore very low dark currents. The persistent photocurrent is due to a very slow return of the nanowire to its neutral state: after illumination, the surface is free of chemisorbed oxygen and the surface depletion is low. The conductivity of the nanowire is high is this state. A positive gate bias helps oxygen becoming trapped at the nanowire surface and restoring the surface depletion faster. The time required to get back to the initial state depends on the dark current requirements for a given application. For a dark current lower than $10^{-13}\, \rm A$, the time is shortened by the gate pulse from hours to a few minutes (see Figure \ref{I_ph_vs_t}), which considerably improves the device operability.  \\

\section{Conclusion}

Single ZnO nanowire photodetectors have been fabricated, and their electrical properties have been investigated in the dark and under illumination. Experiments were performed under various atmospheres and in a temperature range of $100\, \rm K$--$293\, \rm K$. For the first time the existence of a two step photocurrent response was measured and discussed. A high photoconductive gain of $7.8\times 10^{7}$ was determined, associated with very low dark currents, well below $0.1\,\rm pA$, when the device is appropriately biased. These properties demonstrate the potential of nanowires for very high sensitivity photodetection.
The oxygen chemisorption process shows an activation energy of $E_{\rm a} = 115\, \rm meV$.
Finally, an appropriate gate pulse was demonstrated to shorten the persistence of the photocurrent, and restore the photodetector operating condition. This is an important step towards future application of these high gain photodetectors.
Further work is now underway in order to investigate higher rate$/$high gain photodetection operation.

\section{Acknowledgements}
The authors wish to thank the Dr V. Sallet and Dr A. Lusson from the GeMac (University of Versailles) for providing us the nanowires. We acknowledge the financial support of the STSM programme from the COST Action MP1403 'Nanoscale Quantum Optics'. e acknowledge the financial support of the  Labex ACTION (ANR-11-LABX-0001-01).
We acknowledge the financial support of the Grand Est region as all the fabrication was realised using the regional platform “Nanomat” under project “Nanogain”.



\bibliographystyle{iop-num}
\bibliography{references_bibliographique}{}

\end{document}